\documentclass[aps,prb,twocolumn,amssymb,color,pdflatex]{revtex4}

\usepackage{graphicx}
\usepackage{graphics}
\usepackage{dcolumn}
\usepackage{bm}
\usepackage{amsmath}
\usepackage[caption=false]{subfig} 
\usepackage{color}

\usepackage{tikz}

\newcommand{\old}{\color{black}}

\usetikzlibrary{decorations}
\usetikzlibrary{decorations.pathreplacing}
\usetikzlibrary{decorations.markings}

\usepackage{amsmath}

\begin{document}

\newcommand{\bk}{{\bf k}}
\newcommand{\bp}{{\bf p}}
\newcommand{\bv}{{\bf v}}
\newcommand{\bq}{{\bf q}}
\newcommand{\tbq}{\tilde{\bf q}}
\newcommand{\tq}{\tilde{q}}
\newcommand{\bQ}{{\bf Q}}
\newcommand{\br}{{\bf r}}
\newcommand{\bR}{{\bf R}}
\newcommand{\bB}{{\bf B}}
\newcommand{\bA}{{\bf A}}
\newcommand{\ba}{{\bf a}}
\newcommand{\bE}{{\bf E}}
\newcommand{\bj}{{\bf j}}
\newcommand{\bK}{{\bf K}}
\newcommand{\cS}{{\cal S}}
\newcommand{\vd}{{v_\Delta}}
\newcommand{\tr}{{\rm Tr}}
\newcommand{\kslash}{\not\!k}
\newcommand{\qslash}{\not\!q}
\newcommand{\pslash}{\not\!p}
\newcommand{\rslash}{\not\!r}
\newcommand{\bs}{{\bar\sigma}}
\newcommand{\omt}{\tilde{\omega}}
\newcommand{\vv}{\mathcal V}

\newcommand{\qperp}{q_{\perp}}
\newcommand{\qpar}{q_{\parallel}}
\newcommand{\beq}{\begin{equation}}
\newcommand{\eeq}{\end{equation}}

\newcommand{\redtext}[1]{\textcolor{red}{#1}}
\newcommand{\bluetext}[1]{\textcolor{blue}{#1}}

\newcommand{\ket}[1]{| #1 \rangle}
\newcommand{\bra}[1]{\langle #1 |}
\newcommand{\dirac}[2]{\langle #1 | #2 \rangle}

\title{A non-perturbative expression for the transmission through a leaky chiral edge mode }

\author{ Kun W. Kim$^1$, Israel Klich $^2$ and Gil Refael$^1$}
\affiliation{$^1$Department of Physics,
California Institute of Technology, 1200 E. California Blvd, MC114-36,
Pasadena, CA 91125 }
\affiliation{$^2$Department of Physics, University of virginia, Charlottesville, VA 22904 }
\date{\today}
\begin{abstract}
Chiral edge modes of topological insulators and Hall states exhibit non-trivial behavior of conductance in the presence of impurities or additional channels. We will present a simple formula for the conductance through a chiral edge mode coupled to a disordered bulk. For a given coupling matrix between the chiral mode and bulk modes, and a Green function matrix of bulk modes in real space, the renormalized Green function of the chiral mode is expressed in closed form as a ratio of determinants.  We demonstrate the usage of the formula in two systems: i) a 1d wire with random onsite impurity potentials for which we found the disorder averaging is made simpler with the formula, and ii) a quantum Hall fluid with impurities in the bulk for which the phase picked up by the chiral mode due to the scattering with the impurities can be conveniently estimated.  
\end{abstract}
\maketitle

\section{Introduction}

Interest in the behavior of chiral modes on the boundary of 2d insulating systems has been growing in past years. In particular, such modes are always present on the boundaries  of topological states such as topological insulators \cite{kan2005,fu2006,moo2007,fuk2008,qi2008,roy2009,wan2010}, and quantum Hall samples \cite{wen1990,ste2008,tsu1982,von1980}. For example, an important consequence is that zero-temperature electron transport along an edge of a topological insulator will have  a quantized conductance, if time-reversal symmetry is not broken. To describe realistic systems, however, it is always necessary to take into account imperfections, such as potential disorder or impurity scattering \cite{xu2006stability,wu2006helical,cheianov2013mesoscopic}, or even the presence of a bulk states. Interestingly, chiral edge states might even be the result of adding disorder to trivial spin-orbit coupled semiconductors, and the appearance of the so-called Anderson topological insulator \cite{li2009,gro2009,guo2010}. In the course of such a disorder-induced transition, the system necessarily carries extended bulk states as well as chiral modes. Such topological metal systems show a non-trivial behavior of transport properties\cite{ber2010,jun2013}.
 
Disorder effects in non-chiral 1d systems, e.g., localization phenomena due to the introduction of random potential, were also studied theoretically and computationally for many years (for a review see, e.g.  \cite{brandes2003anderson}). While, computationally, it is straightforward to confirm localization of wave functions in 1d single particle systems, the theoretical studies of localization properties and transitions in 1d and higher have proven challenging. A scaling theory suggested in \cite{abr1979}  shows the localization of 1d and 2d at any weak disorder in the system.  For a given distribution of random disorder, an upper bound on localization length was found in \cite{del1983}. Powerful disorder averaging techniques are available, such as the supersymmetry approach \cite{efetov1999supersymmetry}  and the replica trick \cite{weg1979}, which induces an effective non-linear sigma field theory. Also. in strongly disordered systems such as random spin chains, trap models and random polymers,  the role of  quenched disorder on quantum and thermal fluctuations have been studied by the strong disorder renormalization group method (for a review see, e.g. \cite{igl2005}). 

Considerable advances, however, can be made by simple and, arguably, more reliable algebraic methods, when considering a 1d chiral mode connected to a bulk. Such an approach could apply to intrinsic chiral edge modes, and even to localization in non-chiral 1d systems. An example of such a simplification is the subject of the present paper.

In this manuscript, we present a closed-form Green function of a chiral mode coupled to a bulk and its applications. In section II, the Green function is expressed as a ratio of determinants, and involving in the Green function of the bulk and arbitrary couplings between the chiral mode and bulk. An application to disordered strict 1d wire is introduced in section III. The 1d wire is modeled in terms of two chiral modes, and the "det-formula" is applied in the presence of onsite impurities. The disorder averaging of transmission coefficient, the log of transmission coefficient, and the reflection coefficient are worked out in the subsections. In section IV, the phase picked up by the chiral mode in quantum hall fluid system is studied. We conclude in section V. Additional details of disorder averaging of the log of transmission coefficient are worked out in Appendix A. 

\section{Green function of a leaky chiral mode}

\begin{figure}
\centering
\includegraphics[width=80mm]{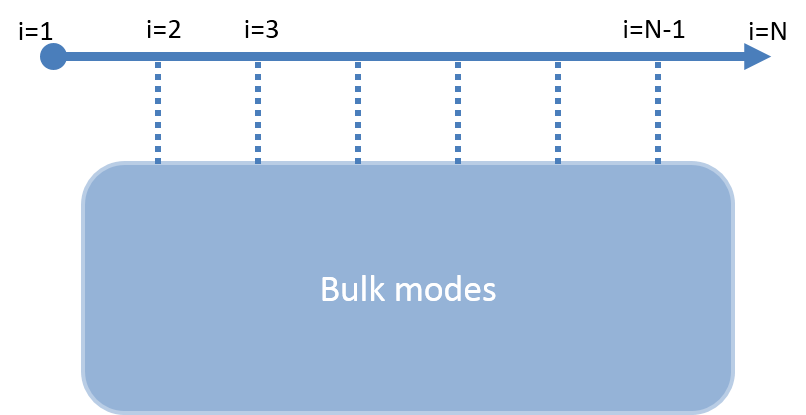}
\caption{A chiral mode is coupled to a bulk that may contain localized and propagating modes. The propagator from site $i=1$ to $i=N$ is found for a given coupling matrix $T$ and Green function of bulk $G_B$ in section II.} \label{fig1}
\end{figure}

In this section, we derive an analytic expression for the chiral Green function between two ends. 

Throughout the paper, we assume that the scattering happens at discrete locations, allowing us to turn the problem into a matrix problem.  We therefore only need the Green function of the right-going chiral mode at the scattering positions:  
\begin{eqnarray}\label{Free Chiral Green}
(G_R)_{nm} = \frac{1}{iv_F} e^{ik|n-m|a} \theta (n-m),
\label{ansatz}
\end{eqnarray}
where $\theta(0)=1/2 $ was introduced. The edge Green function renormalized by the coupling with the bulk modes represented by $G_B$ can be obtained by writing the Hamiltonian of the system in block form and using standard block inversion, with the result: 
\begin{eqnarray}
{\mathcal G}  = \frac{1}{G_R^{-1} - T G_B T^\dagger}.
\label{rightbulk}
\end{eqnarray}
We note that $G_R$, as well as it's inverse $G_R^{-1}$, are lower triangular matrices. For convenience, we pull out phase factors from the Green function, writing it as:  
\begin{eqnarray}
G_R =\hat{U}_t^\dagger G_{R0} \hat{U}_t.
\end{eqnarray}
Here  $\hat{U}_t= e^{-i k a \hat{n}} $, 
where $\hat{n}$ is the position operator along the wire, $\hat{n}_{nm} = \delta_{mn} n$, and $G_{R0}$ is a Green function with zero wave number in Eq.\eqref{Free Chiral Green}.  Due to the particular form of $G_{R0}$, we can write it's inverse explicitly, finding that:
\begin{eqnarray}
G_{R0}^{-1} = -4 v_F^2 \hat{U}_s G_{R0} \hat{U}_s^\dagger,
\end{eqnarray}
where a sign operator  $\hat{U}_s= e^{i \pi \hat{n}} $ was introduced. With these definitions, we now have:
\begin{eqnarray}\label{Green chiral}
{\mathcal G}  = -\frac{1}{4v_F^2} \hat{U}_{st}\frac{1}{G_{R0} + {\mathcal T}   G_B {\mathcal T}^\dagger  }  \hat{U}_{st}^\dagger,
\end{eqnarray}
where $ \hat{U}_{st}= \hat{U}_{s} \hat{U}_{t}$, and ${\mathcal T} =  \hat{U}_{st} T/2v_F $.   

Our main interest in this paper is to get the Green function from one lead to other lead, described by ${\mathcal G}_{N1}$. To do so, we use Cramer's rule, i.e. $C^{-1}_{ij}=M_{ij}det(C)^{-1}$, where $M_{ij}$ is the $ij$'th minor of the matrix $C$, together with Eq.\eqref{Green chiral}, and find: 
\begin{eqnarray}
{\mathcal G}_{N1}  &=& -\frac{e^{i(N-1)ka}}{4v_F^2} \left( \frac{1}{G_{R0} + {\mathcal T}   G_B {\mathcal T}^\dagger  } \right)_{N1} \\
&=& -\frac{e^{i(N-1)ka}}{4v_F^2}  \frac{(-1)^{N+1}M_{1N}}{Det[G_{R0} + {\mathcal T}   G_B {\mathcal T}^\dagger ] } .
\end{eqnarray}
Here:
\begin{eqnarray}
M_{1N} &=& minor_{1N} [G_{R0} + {\mathcal T}   G_B {\mathcal T}^\dagger ] \\
&=& minor_{1N} [-G_{R0}^T + {\mathcal T}   G_B {\mathcal T}^\dagger ] \\
&=& -4 iv_F Det[-G_{R0}^T + {\mathcal T}   G_B {\mathcal T}^\dagger ] \\
&=& -4i v_F (-1)^N Det[G_{R0}^T - {\mathcal T}   G_B {\mathcal T}^\dagger ].
\end{eqnarray}
Here, for the second equality the last row of the matrix $G_0$ is subtracted from all other rows, and for the third equality the two elements at the corner are restored into the determinant. As a result, the renormalized Green function is expressed in terms of the ratio of two determinants: 
\begin{eqnarray}
{\mathcal G}_{N1} 
&=& {\mathcal G}_{N1}^{(0)}  \frac{Det[G_{L0} - {\mathcal T}   G_B {\mathcal T}^\dagger ] }{Det[G_{R0} + {\mathcal T}   G_B {\mathcal T}^\dagger ] } ,
\label{ratio}
\end{eqnarray}
where ${\mathcal G}_{N1}^{(0)}=e^{i(N-1)ka}/iv_F$, and $G_{R0}^T$ is replaced by $G_{L0}$. For the specific system we have in mind in section II, the bulk will be the combination of quantum dots and a left-going chiral mode. 

We will continue the detailed calculation of transmission of disordered 1d wire in the following section.

\section{Example: disordered 1d wire}

 Our first example is a 1d wire with random onsite potentials\cite{and1958}. The physical quantity of interest is the transmission coefficient of the system. Though the system does not contain a chiral mode explicitly, we show in section III.A that the disordered 1d wire can be equivalently modeled in terms of two chiral modes with the couplings through quantum dots. We then apply the det-formula in section III.B to find the transmission coefficient for a given realization of random impurities. In section III.C, the disorder averaging of the transmission coefficient $T$, as well as of $log(T)$, and $1/T$ are computed.

These transport properties of disordered 1d wires have previously been calculated\cite{pen1994, erd1982} in the weak disorder limit with the help of symmetrized products of transfer matrices. The correction order from these methods is significantly limited by the challenge in performing disorder averaging. We present an alternative for disorder averaging using the det-formula from which higher order corrections are easier to access.\old

\subsection{Alternative model of disordered 1d wire}
\begin{figure}
\centering
\includegraphics[width=65mm]{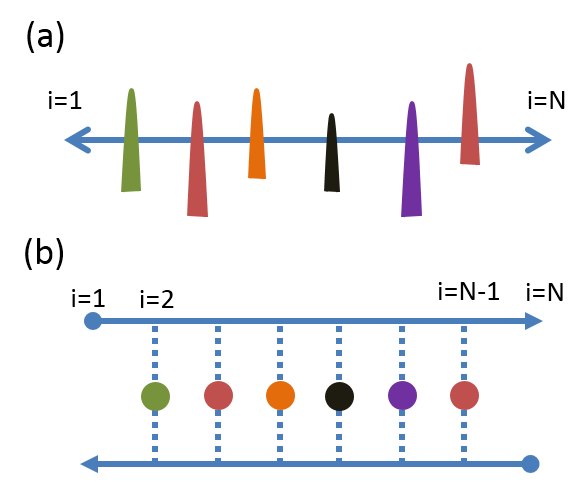}
\caption{(a) 1d wire with random onsite impurities.  (b) An equivalent model of (a). The 1d clean bulk mode is described by two chiral modes, and random impurities are modeled by the coupling between two chiral modes through quantum dots with random onsite potentials. }
\label{fig2}
\end{figure}

 In this section we consider a {\it non}-chiral wire, using the results of the previous section, as a combination of two chiral modes interacting through scatterers. 
The Green function through the disordered 1d wire can be cast in a T-matrix formulation as:
\begin{eqnarray}\label{TmatrixG}
{\mathcal G}_{N1}  = \left( G + G  \frac{1}{I- \alpha G } \alpha G \right)_{N1},
\end{eqnarray}
where $G$ is the free Green function on the wire. We consider a system with $N$ scattering points and placed at intervals of length $a$. Only positions on the wire at which scattering occurs contribute in a non trivial way in Eq.\eqref{TmatrixG}, therefore we can replace the continuous $G$ by an $N\times N$ matrix with elements: 
\begin{eqnarray}
(G)_{nm} = \frac{1}{iv_F} e^{ik|n-m|a}.
\end{eqnarray}
On the other hand, the system includes (N-2) number of impurities with random strengths which can be expressed by a matrix $\alpha$:
\begin{eqnarray}
( \alpha )_{nm} = \delta_{nm} \alpha_n.
\end{eqnarray}
Now consider chiral Green functions with left and right moving modes. One can relate them to the non-chiral Green function by:
\begin{eqnarray}
G = G_R  + G_L.
\end{eqnarray}
where the left-moving chiral Green function is similar to Eq.\eqref{ansatz} with $m$ and $n$ exchanged in the step function, therefore, $(G_L)_{mn}=(G_R)_{nm}$.  Next, we express the Green function through the disordered 1d system, Eq.\eqref{TmatrixG}, in terms of the chiral modes:
\begin{eqnarray}
{\mathcal G}_{N1}  = \left( G_R + G_R  \frac{1}{I- \alpha (G_R+G_L) } \alpha G_R \right)_{N1}.
\label{gn1}
\end{eqnarray}
Note that except for the Green function in the T-matrix, the Green function $G$ on the left and on the right have been replaced by $G_R$, which is justified since $(G)_{nm} = (G_R)_{nm}$  for $n>m$,  while  $\alpha_n$ is nonzero only at the points $N-1>n>1$. 

Finally, we obtain a model of a disordered 1d wire by introducing quantum dots replacing random onsite potentials, and the two chiral modes are connected to the dots through a coupling matrix $T$:
\begin{eqnarray}
\alpha = T G_Q T^\dagger,
\end{eqnarray}
where $G_Q$ is Green function of quantum dots possessing random chemical potentials. By substituting the random potential matrix $\alpha$ in Eq.\eqref{gn1}, now ${\mathcal G}_{N1}$ describes the Green function of right-moving chiral mode as in Fig.2 (b).

\old

\subsection{Green function through disordered 1d wire}
Consider the specific disordered 1d wire system at hand, where the {\it bulk} Green function is:
\begin{eqnarray}
G_{B} = \frac{1}{G_{Q}^{-1} + T^\dagger G_{L}T}
\end{eqnarray}
where the left-going chiral Green function renormalizes the Green function of quantum dots. Note the hopping matrix $T$ has now a different order compared to Eq.\eqref{rightbulk}. For our purpose, it is convenient to consider with a simplified form: 
\begin{eqnarray}
G_L = -\frac{1}{4v_F^2}  \hat{U}_{st}^\dagger G_{L0}^{-1}  \hat{U}_{st}
\end{eqnarray} 
Define ${\mathcal T}_2 =  \hat{U}_{st}^\dagger T/2v_F $,  plug in the left-going chiral Green function back to the bulk Green function expression: 

\begin{eqnarray}
G_{B} = \frac{1}{G_{Q}^{-1} + {\mathcal T}_2^\dagger G_{L0}^{-1}{\mathcal T}_2}
\end{eqnarray}
Next, let us work with the renormalized Green function for this bulk system. Eq.\eqref{ratio}  is reduced to:
\begin{eqnarray}
\frac{{\mathcal G}_{N1}}{ {\mathcal G}_{N1}^{(0)} }
&= &  \frac{Det[G_{L0} - {\mathcal T}  G_B {\mathcal T}^\dagger ] }{Det[G_{R0} + {\mathcal T}   G_B {\mathcal T}^\dagger ] } \\
&= &  \frac{det[G_{L0}({\mathcal T}   G_B {\mathcal T}^\dagger)^{-1} - I ] }{det[G_{R0} ({\mathcal T}   G_B {\mathcal T}^\dagger)^{-1}+I ] } \\
&= &  \frac{Det[G_{L0} -(I -G_{L0}\hat{U}_{t}^2 G_{L0}^{-1}\hat{U}_{t}^{\dagger 2}) {\mathcal T} G_{Q} {\mathcal T} ^\dagger] }{Det[G_{R0} +(I +G_{R0} \hat{U}_{t}^2 G_{L0}^{-1}\hat{U}_{t}^{\dagger 2}) {\mathcal T} G_{Q} {\mathcal T} ^\dagger] } \\
&= &  \frac{Det[G'_{L0} -R {\mathcal T} G_{Q} {\mathcal T} ^\dagger] }{Det[G'_{R0} + {\mathcal T} G_{Q} {\mathcal T} ^\dagger] }
\end{eqnarray}
where in the second equality the determinant with lower case is designated as the determinant of matrix excluding the boundary elements. Next, the inverse of ${\mathcal T}  G_B {\mathcal T}^\dagger$ is expressed using the relation $ {\mathcal T}_2 {\mathcal T}^{-1} =\hat{U}_{t}^{\dagger 2}= e^{i 2ka \hat{n}}$  and ${\mathcal T} G_{Q} {\mathcal T} ^\dagger$ recovering a determinant of a whole matrix. In the last equality, new Green functions have been defined:
\begin{eqnarray}
G'_{R0} &=& \frac{1}{G_{R0}^{-1}+\hat{U}_{t}^{ 2} G_{L0}^{-1}\hat{U}_{t}^{\dagger 2}} \\
&=& -\frac{1}{4v_F^2} \hat{U}_{st}  \frac{1}{ G_{R}  +  G_{L} } \hat{U}_{st}^{\dagger}
\end{eqnarray}
where $G_R + G_L$ is the Green function of a clean 1d wire Hamiltonian with nearest neighbor hopping, therefore the inverse is a tridiagonal matrix. On the other hand, 
\begin{eqnarray}
G'_{L0} &=& \frac{1}{I+G_{R0} \hat{U}_{t}^2 G_{L0}^{-1}\hat{U}_{t}^{\dagger 2}}   G_{L0} \\
&=& G'_{R0} G_{R0}^{-1}G_{L0}
\end{eqnarray}
and 
\begin{eqnarray}
R &=& \frac{1}{I+G_{R0} \hat{U}_{t}^{ 2} G_{L0}^{-1} \hat{U}_{t}^{\dagger 2}}   (I+G_{L0} \hat{U}_{t}^{ 2} G_{L0}^{-1} \hat{U}_{t}^{\dagger 2}) \\
&=& I - G'_{R0} (I+G_{R0}^{-1} G_{L0}) \hat{U}_{t}^{ 2} G_{L0}^{-1} \hat{U}_{t}^{\dagger 2}. \label{R}
\end{eqnarray}
Studying the second term on the right side of  Eq.\eqref{R},  equal to $(I-R)$, we find that, $(I-R)_{nm}=0$ for $n>1$, and $(I-R)_{11}=1$.  As a result, we find that $(R)_{11}=0$ and $(R)_{nn}=1$ for $n\neq 1$. Finally, one can verify that: 
\begin{eqnarray}
(G'_{L0} +RG'_{R0} )_{nm} = \frac{1}{4i} \delta_{n,1} \delta_{m,N},
\end{eqnarray}
and the renormalized Green function is reduced to:
\begin{eqnarray}\nonumber
\frac{{\mathcal G}_{N1}}{{\mathcal G}_{N1}^{(0)}} &= &  Det \left[ (G'_{L0} +RG'_{R0} ) \big/  (G'_{R0} + {\mathcal T} G_{Q} {\mathcal T} ^\dagger ) - R \right] \\ \nonumber
&=& \frac{1}{Det[I + {\mathcal T} G_{Q} {\mathcal T} ^\dagger G^{'-1}_{R0}]  }\\ \label{1ddet}
&=& \frac{1}{Det[I - \alpha(G_R+G_L)] } \label{eq35}
\end{eqnarray}
using the notation previously defined in the alternative model.  
A Green function of a similar form was discovered by Thouless\cite{tho1972},  and used to obtain the density of state. Once the transmission of a 1d disordered system for particular disorder realization is written in this way, the analytic manipulation becomes easier and the disorder averaging is rendered accessible as we present in the next sections. \old

\subsection{Disorder averaging}
 Carrying out disorder averaging has been a tricky problem even for strict 1d wire\cite{erd1982,pen1994}. In this section with the help of the det-formula, Eq.\eqref{1ddet}, the disorder averaging of several different quantities related to the transport of a disordered strict 1d wire are simplified and compared with numerical disorder averaging results. 

\subsubsection{Transmission coefficient:  $\overline{T}$}

In the weak disorder limit we make use of the Green function Eq.\eqref{1ddet} to obtain the transmission coefficient through the disordered wire, $T=|{\mathcal G}_{N1}|^2$. The determinant can be perturbatively expanded using $det(I+O) = exp(Tr(O - O^2/2 + \cdots )$. In our case, the matrix $O$ is simply $-\alpha (G_R+G_L)$. We now consider the traces:
\begin{eqnarray}
(O)_{nn} &=& - \frac{\alpha_n}{iv_F} \\
(O^2)_{nn} &=& - \sum_m \frac{\alpha_n \alpha_m }{v_F^2}e^{2ika|n-m|}.\label{eqo2}
\end{eqnarray}
We see that the diagonal elements of $O$ are purely imaginary, therefore the $Tr O$ term just provides a phase to the Green function and the transmission coefficient is unchanged by the first order term. In contrast, the diagonal elements of $O^2$ possess a non-vanishing real part, and therefore the leading order contribution in the weak impurity limit.

\begin{figure}
\centering
\includegraphics[width=85mm]{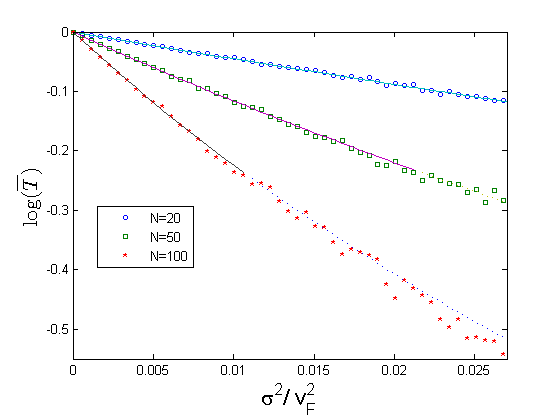} 
\caption{  $\log \overline{T}$ of a disordered 1d wire with $ka=0.45\pi$ is plotted. 1000 different disorder realizations are  numerically performed (dots) and compared with the analytic expression within the perturbative regime (solid lines) and non-perturbative regime (dashed lines) for different system size N=20, 50, 100. The perturbative regime is determined by $\lambda = \sigma^2/v_F^2 < 1/N$ according to Eq.\eqref{condition}. }
\label{fig:fig3}
\end{figure}

For a given realization of impurity strengths $\{\alpha \}$, we have the transmission coefficient:
\begin{eqnarray}
T(\{\alpha \}) &=& e^{ -{Tr \left[ \sum_{m=1}^{M_{up}} \frac{(-1)^{m+1}}{m} (O^m+O^{*m})\right] } } \label{eqT}, \\
 &\simeq& \frac{1}{exp(-Tr(O^2+O^{*2})/2)}, \\
&=& \frac{1}{exp ( \sum_{n,m} \alpha_n \alpha_m cos (2ka|n-m|)/v_F^2)},
\end{eqnarray}
 where the upper bound of sum in the exponent of Eq.\eqref{eqT}, $M_{up}=2$ is taken in the second equality.  The indices $n$ and $m$ run over the lattice positions between $n=2$ to $n=N-1$. The bilinear summation over impurity strengths is particularly useful for analytic disorder averaging if the $\alpha_i$ are given by a Gaussian distribution:
\begin{eqnarray}
\overline{T} &=& \int T(\{\alpha \})  \prod_{i=2}^{N-1} \frac{e^{- \alpha_i^2/2\sigma^2}}{\sqrt{2\pi} \sigma} d\alpha_{i}  \\
&=& \frac{1}{\sqrt{det(K)}}
\end{eqnarray}
where the matrix K is:
\begin{eqnarray}
(K)_{nm}&=&\delta_{nm} +2 \sigma^2 cos(2ka|n-m| )/v_F^2 .
\end{eqnarray}
$K$ can also be written in the form
\begin{eqnarray}
K &=&  I_{N-2} + \frac{2\sigma^2}{v_F^2}(|v\rangle\langle v| + |v^*\rangle\langle v^*|),
\end{eqnarray}
where $|v\rangle$ is a vector with elements $(v)_n= e^{inka}$ with the index $n=1,\cdots,N-2$. 
This form shows that $K$ is involves a rank 2 matrix, and the determinant of K is as easy as taking a determinant of a $2\times2$ matrix. As a result:
\begin{eqnarray}
det(K) = \left[ 1-(N-2)\frac{\sigma^2}{v_F^2}\right] ^2 - \frac{\sigma^4}{v_F^4}|\Sigma_{n=1}^{N-2} e^{inka}|^2.
\end{eqnarray}
The second term in the above equation is important when the Fermi velocity $v_F=-2 ta sin(ka)$ is near zero. Otherwise, the first term gives the dominant contribution to the transmission coefficient.   Fig.\ref{fig:fig3} shows  a surprisingly good agreement between numerics that take into account all orders and the analytics within 2nd order for different system size N. The perturbation theory works only when the expansion parameter is smaller than unity. For our case, from the expansion parameter, $Tr(O^2)$ in Eq.\eqref{eqo2}, the condition is
\begin{eqnarray}
\lambda = \frac{\sigma^2 N} {v_F^2}<1.
\label{condition}
\end{eqnarray}
The horizontal axis of the plot is $\sigma^2/v_F^2$, and therefore the analytic expression is valid for $\sigma^2/ v_F^2<1/N$. For the largest system size $N=100$ in the plot, we can see the deviation as early as $\sigma^2 /v_F^2\simeq 0.01$, consistent with the discussion. 

\begin{figure}[h]
\centering
\includegraphics[width=85mm]{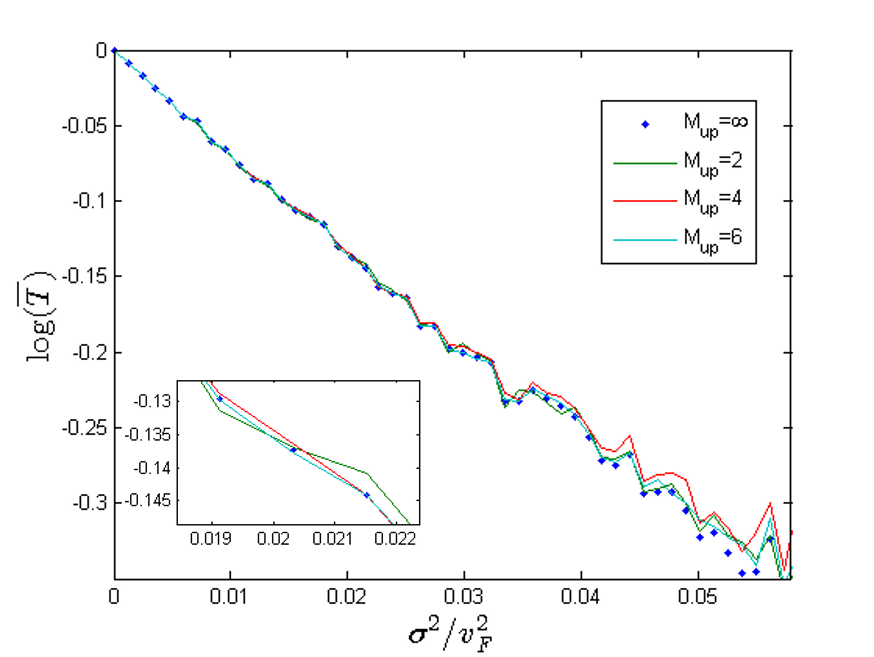}
\caption{ $\log \overline{T}$ with $ka=0.45\pi $ is numerically computed (lines) considering different upper bound $M_{up}$ of summation in the exponent of Eq.\eqref{eqT}, and compared with exact numerical results (dots) for a system of  $N=30$. The improvement by including the higher order correction is slight.   }
\label{fig4}
\end{figure}

A natural question is: how will the estimates improve if higher order terms in the matrix $O$ are taken into account. We found that the improvement is negligibly small as shown in Fig.\ref{fig4} that shows the comparison for the inclusion of different orders. This is an interesting point to discuss. The random variables, i.e. the strengths of the impurities, $\alpha_n$, appear in the exponent of exponential function. Therefore, the average transmission coefficient is dominated by the realizations with very small impurity strengths, the contribution of other realizations are exponentially suppressed. Because at small impurity strengths the contribution of higher order terms in Eq.\eqref{eqT} are also small, we find that the leading order actually gives a good estimate of all orders. 

\subsubsection{Localization length: $\overline{log (T)}$}

As a reference to actual experiments, the average of the transmission coefficient is not a good measure, since it is dominated by realizations with a small probability. Instead, the quantity considered in this section, namely the average of $log(T)$, is a more reliable quantity.

The direct calculation of the average $log(T)$ has been limited to the first order so far because of the difficulty \cite{pen1994} to deal with the magnitude of disorder averaged Green functions or the product of transfer matrices. Instead, authors employed an asymptotic relation to make use of the self averaging property of the quantity \cite{hal1965}. In this section, we explicitly compute the disorder averaged $log(T)$ in the weak and strong disorder strength limit with the help of the det-formula. 

Let us first consider the weak disorder strength case, $W/v_F <1$, where the $W$ is cutoff of uniform disorder distribution. Here we employ the uniform distribution instead of a Gaussian distribution, ensuring all moments are bounded. This weak disorder limit validates the expansion of $log(I+O)$ as previously done in the exponent of Eq.\eqref{eqT}. We consider the even terms in the expansion, which are the only non-vanishing terms after averaging, with the $m^{th}$ term given by:
\begin{eqnarray}
\frac{1}{m}Tr\left[O^{m}+O^{*m} \right] = \\
\frac{2}{m}\sum_{n_1,\cdots ,n_{m}} \left(\prod_{i=1}^{m} \frac{\alpha_{n_i}}{v_F}\right) cos\left( ka \sum_{i=1}^{m} |n_{i+1} - n_i|\right)
\label{indices}
\end{eqnarray}
where $n_{m+1} = n_1$. 

\begin{figure}[h]
\begin{center}
\includegraphics[width=85mm]{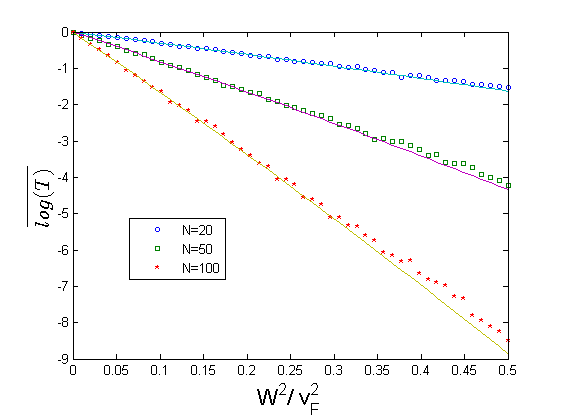}
\caption{  $\overline{log(T)}$  for different system sizes, N=20, 50, 100 with $ka=0.25\pi$ using uniform distribution $[-W,W]$ is numerically computed (dots) and compared with the analytic results (lines), Eq.\eqref{logTave}. Despite ignoring multiple scatterings from different impurities, as a leading order approximation the analytic result works well beyond its pertrubative regime $W^2/v_F^2 < 1/N$. The disorder averaging with 1000 realizations is performed.  }
\label{fig5a}
\end{center}
\end{figure}

The exact summation over indices cannot be performed exactly, however the dominant contribution comes from the terms with zero argument of the cosine function: $n_{i+1}=n_i$.  Retaining only such terms, we find that the average of the log transmission is given by:\begin{eqnarray}
\overline{log(T)} &\simeq & -\frac{L}{a}\sum_{m=1}^\infty \int_{-W}^W (-1)^{m}\frac{2}{m} \left( \frac{\alpha}{v_F} \right) ^m  \frac{d\alpha }{2W} \\
 &=&-\frac{L}{a} \sum_{m=1}^\infty \frac{1}{m(2m+1)} \left( \frac{W}{v_F}\right)^{2m} \\
&=& -L/l_{loc}.\label{logTave}
\end{eqnarray}
 where $L=(N-2)a$ is the length of disordered regime.  The next order correction with non-zero argument of the cosine function is discussed in the appendix.  Here the localization of the system is found as:
\begin{eqnarray}
l_{loc}/a =  \left[ Log\left(1-\frac{W^2}{v_F^2}\right) -\frac{v_F}{W}log \left( \frac{1+W/v_F}{1-W/v_F} \right) -2 \right]^{-1}, \label{loc}
\end{eqnarray}
which is the analytic expression and plotted in Fig.\ref{fig5a} compared with computational results that takes into account all non-zero arguments of cosine function.  

\begin{figure}[h]
\begin{center}
\includegraphics[width=85mm]{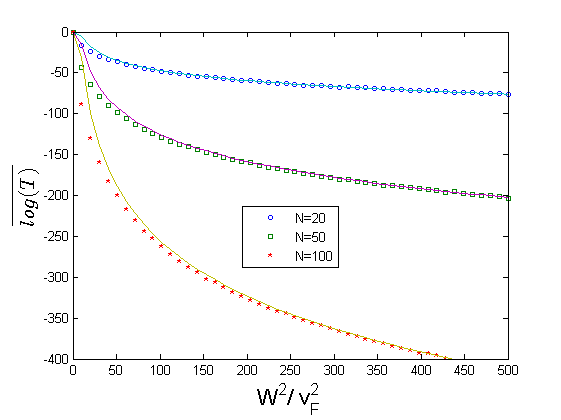}
\caption{  $\overline{log(T)}$ in strong disorder strength limit with $ka=0.25\pi$ is numerically computed (dots) and compared with analytic approximation (lines) according to Eq.\eqref{strong}. They matches  well in the strong disorder regime, while in the weak disorder regime the deviation is present since the perturbation parameter is now $\lambda = v_F^2/W^2$, and the range of impurity strength is  $[-W,W]$ instead of $[-1/W,1/W]$. The disorder averaging with 1000 realizations is performed. }
\label{fig5b}
\end{center}
\end{figure}

For strong disorder strength limit, the perturbative expansion done above does not work. Instead, we can perform the perturbation of the matrix $1/\alpha G$:

\begin{eqnarray}
T
&=& \left| \frac{1}{det[I + \alpha G ] } \right|^2  \simeq   \left| \frac{1}{det[\alpha G]} \right|^2 
\end{eqnarray}
\begin{eqnarray}
log(T) =C + \sum_{i=2}^{N-1} log\left(\frac{v_F^2}{\alpha_i^2} \right)
\end{eqnarray}
where $C = -2log|det[v_F G]|$. Therefore, the average of the log g in the uniform distribution of disorder comes straightforward:

\begin{eqnarray}
\overline{log(T)} =C-2(N-2)\left[ log\left( \frac{W}{v_F}\right)-1 \right], \label{strong}
\end{eqnarray}
which is plotted in the Fig.\ref{fig5b} for different wave numbers and shows good agreement.

\subsubsection{Resistence: $\overline{R}=\overline{1/T} $}

Finally, in this section we discuss the average of $R=1/T$, the reflection coefficient of system, the inverse of transmission coefficient before disorder averaging. Although the reflection coefficient may not be a physically relevant quantity, there has been interest to compute it analytically upon disorder averaging\cite{pen1994}. Here,  we do not have an  analytic expression and we suggest a different approach. Directly from Eq.\eqref{eq35} the reflection coefficient can be expressed by the determinant of the matrices:
\begin{eqnarray}
1/T &=&\frac{1}{ {\mathcal G}_{N1} {\mathcal G}_{N1}^*}\\
&=& \frac{det[G^{-1} G^{*-1}-\alpha G^{*-1} - G^{-1} \alpha +\alpha \alpha]} {det[G^{-1} G^{*-1}]},\label{Reflection}
\end{eqnarray}
with $G^{-1} = 1/(G_R+G_L)$, a tridiagonal matrix, as before. Therefore, the matrix inside the determinant  in the   numerator in Eq.\eqref{Reflection} is essentially an extended version of a tridiagonal matrix which has five non-zero elements along the diagonal element instead of three as in Eq.\eqref{eq35}, with the elements:
\begin{eqnarray}
D_{n,n} &=& (\alpha_n-E)^2 + 2 t^2 \\
D_{n+1,n} &=& D_{n,n+1} = t(\alpha_n + \alpha_{n+1})-2tE \\
D_{n+2,n} &=& D_{n,n+2}= t^2 .
\end{eqnarray}
Here $D_{1,1} =D_{N,N}= 2t^2$, $\alpha_1=\alpha_N=te^{-ika}$, and we  place impurities at $i=6,10,14,\cdots,N-5$ so that each transfer matrix is not correlated. As a result the format of the transfer matrix can be written using 6 elements: 
\begin{eqnarray}
\begin{pmatrix} d_{4(m+1)+1} \\  d_{4(m+1)+2} \\ d_{4(m+1)+3} \\ d_{4(m+1)+4} \\ X_{4(m+1)+2} \\ X_{4(m+1)+3} \end{pmatrix} = T^m_{6\times 6}\label{deteq1}
\begin{pmatrix}  d_{4m+1} \\ d_{4m+2} \\ d_{4m+3} \\ d_{4m+4} \\ X_{4m+2} \\ X_{4m+3} \end{pmatrix} \nonumber
\end{eqnarray}
where $d_{4m+j}$ is the determinant of the upper left square matrix of D up to the $(4m+1)^{th}$ row and column. $X_n$ are also the determinants of a similar matrix which is required to construct the transfer matrix:
\begin{figure}[h]
\begin{center}
\includegraphics[width=85mm]{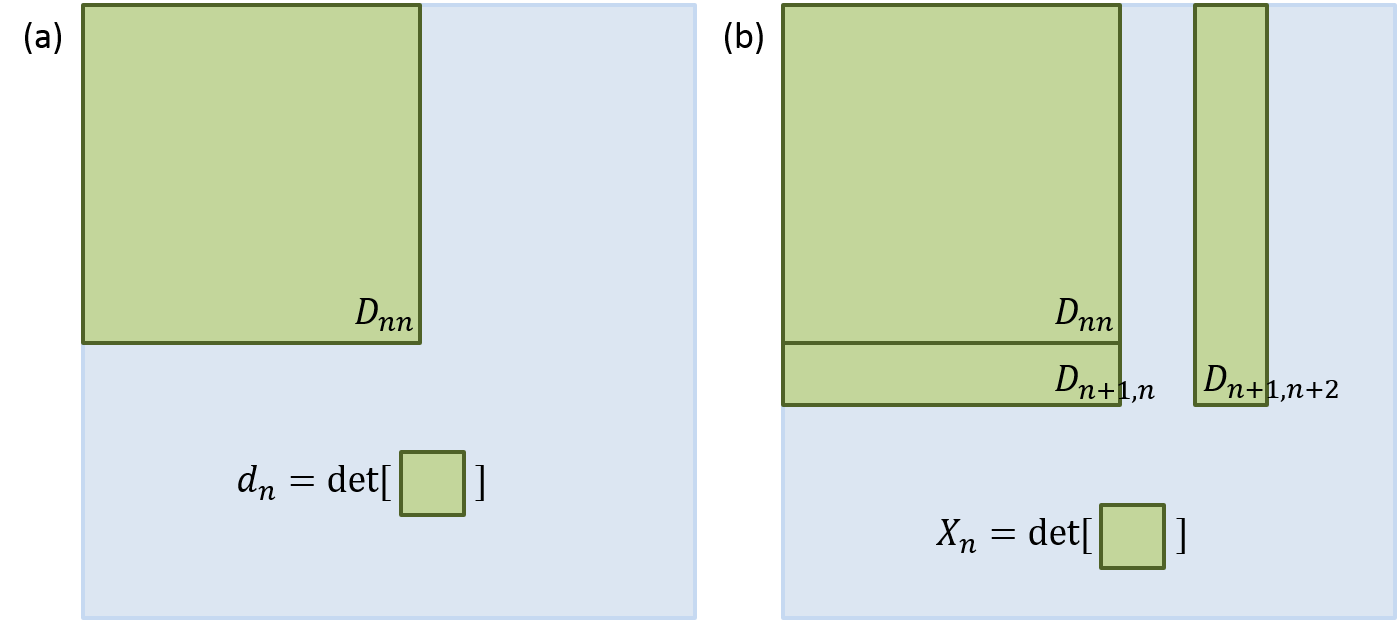}
\caption{Describing the elements of the determinant in Eq.\eqref{deteq1}. (a) $d_n$ is the determinant of Green square block up to element $D_{nn}$. (b) $X_n$ is the determinant of the Green blocks collection.   }
\end{center}
\end{figure}

\begin{eqnarray}
X_{n} = D_{n,n+1}d_{n-1}- D_{n-1,n+1}D_{n,n-1}d_{n-2} \\
-D_{n-1,n+1} D_{n-2,n}X_{n-2}
\end{eqnarray}
From these relations one can build the train of transfer matrices and perform disorder averaging for each matrix separately. 

\begin{figure}[h]
\begin{center}
\includegraphics[width=85mm]{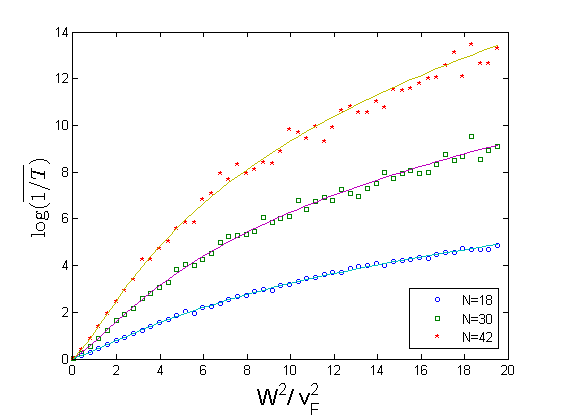}
\caption{  $\log\overline{1/T}$ for different system sizes, N=18, 30, and 42 with $ka=0.45\pi$ is computed numerically (dots) by disorder averaging of 2000 realization. And, the analytic disorder averaging (lines) is performed exactly using transfer matrix, though its eigenvalues cannot be expressed in a closed form. }  \label{fig8a}
\end{center}
\end{figure}

As an analytic result, we look for the exponents of the $1/R$, which are related to  the eigenvalues of transfer matrix. The characteristic equation of $T^m_{6\times 6}$ has one trivial solution, unity, and the other solutions satisfy a quintic equation whose solution is not analytically available.  Here we show the computation of "analytic" and numeric results in Fig.\ref{fig8a}. The plot shows the exponential increase of the resistance of disordered wire with the strength of impurities. 

\section{Example: Qauntum Hall fluid}

\begin{figure}[h]
\begin{center}
\includegraphics[width=85mm]{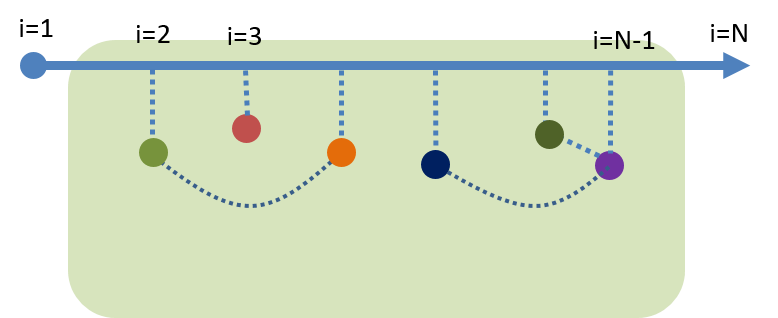}
\caption{ A chiral mode in quantum Hall system with an insulating bulk. Multiple impurities with random couplings near the edge are encountered  by the chiral mode, and the accumulated phase is estimated with the det-formula. }\label{fig8}
\end{center}
\end{figure}

Another natural application for the det-formula is the quantum Hall system.  The application of the det-formula is rather straightforward as it contains an explicit chiral model along the boundary. Consider the system described in Fig.\ref{fig8} where impurities near the boundary may scatter the chiral mode, and  impurities are also connected with each other in a random fashion. In the single particle picture, the conductance of the chiral mode is expected to be quantize, however, the Green function will possess a system specific phase factor due to the scattering from the impurities. These non-trivial phase factors in quantum Hall systems can be probed by different types of interferometers\cite{cha1997,fel2006,fel2007,akh2009,bis2008}. Our purpose in this section is to estimate the accumulated phase throughout the system. 

Let us introduce the Green function, $G_B$, of quantum dots with random coupling, the coupling matrix $T$ between chiral mode and quantum dots, and $\mathcal T=\hat{U}_{st} T/2v_F$, as done in section II.  The direct application of the det-formula gives:
\begin{eqnarray}
{\mathcal G}_{N1} 
&=& \frac{e^{i(N-1)ka}}{iv_F}  \frac{det[I- G_{L0} ^{-1} {\mathcal T}   G_B {\mathcal T}^\dagger ] }{det[I+ {\mathcal T}   G_B {\mathcal T}^\dagger G_{R0} ^{-1} ] } \\
&=& \frac{e^{i(N-1)ka}}{iv_F}  \frac{e^{Tr Log[I-G_{L0} ^{-1} {\mathcal T}   G_B {\mathcal T}^\dagger ]} }{e^{TrLog[I- (  G_{L0} ^{-1}  {\mathcal T}    G_B^\dagger  {\mathcal T}^\dagger)^{\dagger}]} },
\end{eqnarray}
where in the first equality the order of matrix multiplication is changed within the determinant of denominator, then in the second equality $(G_{R0} ^{-1})^{\dagger}=-G_{L0} ^{-1}$ is used. Having a symmetric form,  diagonalization helps to further simplify:
\begin{eqnarray}
G_{L0} ^{-1}  {\mathcal T}   G_B {\mathcal T}^\dagger = U^{-1}  \Lambda U
\end{eqnarray}
The bulk Green function $G_B$ may, in general, not be a hermitian matrix in an open, lossy, system. However, when the bulk states are insulating and do not contain propagating states to the exterior, the bulk Green function becomes real and symmetric, $G_B^\dagger = G_B$. As a result, the renormalized Green function is now reduced to: 
\begin{eqnarray}
{\mathcal G}_{N1} 
&=& \frac{e^{i(N-1)ka}}{iv_F} e^{\sum_n Log\left[  \frac{1-\lambda_n}{1- \lambda_n^*}  \right]} 
\end{eqnarray}
which shows how the chiral mode electron picks up phases throughout a system with impurities. Given the distribution of eigenvalues of the matrix $ {\mathcal T}   G_B {\mathcal T}^\dagger$,  it would also be possible to derive the distribution of phases. 

\section{Conclusion}

In this manuscript we introduced a closed form expression for the propagator of chiral mode coupled to a bulk. For a given bulk Green function and coupling matrix to the chiral mode, we were able to express the renormalized chiral Green function by the ratio of determinants. No assumption is made on the bulk Green function, therefore, the formula is applicable to a system with and without non-localized bulk modes.

 Using a disordered 1d quantum wire as example, we demonstrated how non-chiral 1d systems can be modeled in terms of chiral modes, and computed the average transmission coefficient, its inverse, and log. The det-formula was shown to be especially powerful in  performing  disorder averaging for different transport quantities. A similar trick is expected to work for quasi 1d and topological metal systems. 

As a second example, the phase accumulated by the chiral mode is conveniently estimated in a quantum hall system with random impurities. This formula may be used to describe the interference pattern in quantum hall interferometers as a function of chemical potential of the bulk and other system characteristics. 

It is a pleasure to acknowledge useful discussions with Matthew Hastings and Konstantin Efetov as well as funding from the IQIM, an NSF center, supported by the Moore foundation, and from DARPA through FENA. IK acknowledges financial support from NSF CAREER award No. DMR-0956053.

\section{Appendix}

\subsection{ Localization length: next order}

In section III.C.2, we obtained the leading order of disorder averaged $log(T)$, assuming all indices in Eq.\eqref{indices}  are the same: $n_1=n_2=\cdots =n_m$. For the next order calculation, in this appendix we allow the indices to be two different indices: $n_1, \cdots, n_m \in n_i, n_j$ with $i\neq j$. Then all possible terms of order m can be expressed by the help of a $2\times 2$ transfer matrix form:
\begin{eqnarray}&
\sum_{n_1,\cdots,n_m}\left(\prod_{i=1}^{m} \frac{\alpha_{n_i}}{v_F}\right) Re[e^{i ka \sum_{i=1}^{m} |n_{i+1} - n_i|}], \\ \nonumber &
= \sum_{n_1,n_2}Re Tr \left[
\begin{pmatrix}
\alpha_{n_1} & \alpha_{n_1} e^{ik\Delta} \\
\alpha_{n_2} e^{-ik\Delta} & \alpha_{n_2}  
\end{pmatrix}
^m- 
\begin{pmatrix}
\alpha_{n_1}^m & 0 \\
0 & \alpha_{n_2} ^m
\end{pmatrix}
\right],
\end{eqnarray}
 where $\Delta= |n_1-n_2|a$ is the distance between two impurities.  Now the problem is reduced to finding eigenvalue of the transfer matrix:
\begin{eqnarray}
\lambda_{\pm} = \frac{\alpha_{n_1}+\alpha_{n_2}}{2} \pm \sqrt{\frac{(\alpha_{n_1}-\alpha_{n_2})^2}{4} + \alpha_{n_1}\alpha_{n_2}e^{2ik\Delta}}
\end{eqnarray}
By using the prefactor in Eq.\eqref{indices} and inserting the eigenvalues:
\begin{eqnarray}
log(T_{2})&=&\sum_{n_1,n_2} \frac{(-1)^m}{mv_F^m}2 Re[ \lambda_+^m+\lambda_-^m-\alpha_{n_1}^m-\alpha_{n_2}^m] \\
&=&2Re\sum_{n_1,n_2} log\left[\frac{(1+\frac{\lambda_+}{v_F})(1+\frac{\lambda_-}{v_F})}{(1+\frac{\alpha_{n_1}}{v_F})(1+\frac{\alpha_{n_2}}{v_F})}\right]\\
&=&2Re\sum_{n_1,n_2} log\left[1-\frac{\alpha_{n_1}\alpha_{n_2}e^{2ik\Delta}/v_F^2}{(1+\frac{\alpha_{n_1}}{v_F})(1+\frac{\alpha_{n_2}}{v_F})}\right]
\end{eqnarray}
which is the generating function for the correction of two indices to all orders. For example, the first correction appears in the fourth order with non-zero argument of cosine function. By expanding the log in series of $\alpha_i/v_F$, then disorder averaging:

\begin{eqnarray}
\overline{log(T_2^{(4)})}&=&Re\sum_{n_1,n_2} \frac{\overline{\alpha_{n_1}^2}\overline{\alpha_{n_2}^2}(4e^{2ik\Delta}+2e^{4ik\Delta})}{-2v_F^2}  \nonumber \\
&=&\frac{\left< \alpha^2 \right>^2}{-v_F^4}\sum_{\Delta/a=1}^{L/a-1}\frac{L-\Delta}{a}[2cos(2k\Delta )+cos(4k\Delta)] \nonumber \\ 
&=&\frac{\left< \alpha^2 \right>^2}{v_F^4} \left[\frac{3 L}{2a} - \frac{sin^2(kL)}{sin^2(ka)}-\frac{sin^2(2kL)}{2sin^2(2ka)}\right]
\end{eqnarray} 
 where $L=(N-2)a$ is the length of disordered regime.  One can see that the term in the square bracket may not be small depending on the wavenumber k. But, the term goes to zero upon the integration over k (see the Fig.9). This is the case for all higher order corrections. By carefully performing the expansion of the generating function, the general expression for $m^{th}$ order correction of two indices can be deduced:

\begin{figure}[h]
\begin{center}
\includegraphics[width=75mm]{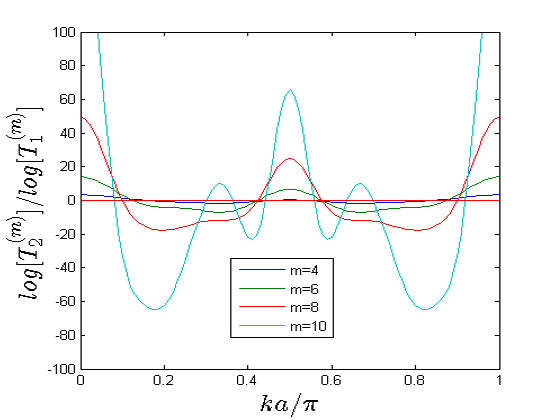}
\caption{ The ratio of the single-index approximation $log[T_1^{(m)}]$,  Eq.\eqref{logTave}, and the double-index correction $log[T_2^{(m)}]$, Eq.\eqref{aveTm},  for different orders of expansion, $m$, as function of the wave number $k$. At most of wave numbers the ratio exceeds unity and this leads the stronger fluctuation of the transmission coefficient for the higher expansion order terms as shown in Fig.\ref{fig4} as approaching the perturbation limit.  }\label{ratiologTcorrection}
\end{center}
\end{figure}

\begin{eqnarray}\label{aveTm}
\overline{log(T_2^{(m)})}=\frac{1}{m}\sum_{n=2}^{m-2}\sum_{l=1}^{l_{up}}\frac{1}{n+1}\frac{1}{m-n+1}\left(\frac{W}{v_F}\right)^m \nonumber \\
 \times{m-n\choose l} {n \choose l} l \left( \frac{1}{n} +\frac{1}{m-n}\right) \left[ L-\frac{sin^2(kLl)}{sin^2(kal)} \right]
\end{eqnarray}
where $l_{up}=min(n,m-n)$. Again, the term in the square bracket is on average of wavenumber $k$. In Fig. \ref{ratiologTcorrection} the ratio between the correction and the leading order term computed in the main text is plotted for different orders m. The fluctuation with wavenumber is larger for higher order terms. Nevertheless, within the perturbation limit, those large fluctuations are not noticeable because they are suppressed by $(W/v_F)^{m}$. However, this fluctuation explains why the inclusion of higher order results in the poor behavior near the non-perturbative regime, observed in Fig. \ref{fig4}.

\bibliographystyle{apsrev}
\bibliography{DisorderRefs_kunz}

\end{document}